\documentclass[11pt,a4paper]{article}
\usepackage{amsmath,amssymb}
\usepackage{epsfig,graphicx}

\begin{document} 

\title{\bf Highly sensitive gamma-spectrometers of GERDA for material screening: Part I.}
\author{\underline{D.~Bud}j\underline{\'a\v{s}}$^a$\footnote{{\bf e-mail}: dusan.budjas@mpi-hd.mpg.de}, C. Cattadori$^b$, A. Gangapshev$^c$, W. Hampel$^a$, M. Heisel$^a$,\\ G. Heusser$^a$, M. Hult$^d$, A. Klimenko$^{c,e}$, V. Kuzminov$^c$, M. Laubenstein$^b$,\\ W. Maneschg$^a$, S. Nisi$^b$, S. Sch\"onert$^a$, H. Simgen$^a$, A. Smolnikov$^{c,e}$,\\ C. Tomei$^b$, A. di Vacri$^b$, S. Vasiliev$^{c,e}$, G. Zuzel$^a$\\
$^a$\small{\em Max-Planck-Institut f\"ur Kernphysik, Saupfercheckweg 1, 69117 Heidelberg, Germany}\\
$^b$\small{\em Laboratori Nazionali Del Gran Sasso, SS 17 bis Km18+910, 67010 Assergi (AQ), Italy} \\
$^c$\small{\em Joint Institute for Nuclear Research, 141980 Dubna, Russia}\\
$^d$\small{\em EC-JRC-Institute for Reference Materials and Measurements, Retieseweg 111, 2440 Geel, Belgium} \\
$^e$\small{\em Institute for Nuclear Research, 117312 Moscow, Russia}\\
}
\date{October 5, 2007}

\maketitle
\begin{abstract}

The GERDA experiment aims to search for the neutrinoless double beta-decay of $^{76}$Ge and possibly for other rare processes. The sensitivity of the first phase is envisioned to be more than one order of magnitude better than in previous 0$\nu \beta \beta$-decay experiments. This implies that materials with ultra-low radioactive contamination need to be used for the construction of the detector and its shielding. Therefore the requirements on material screening include high-sensitivity low-background detection techniques and long measurement times.

In this article, an overview of material-screening laboratories available to the GERDA collaboration is given, with emphasis on the gamma-spectrometry. Additionally, results of an intercomparison of the evaluation accuracy in these laboratories are presented.
\end{abstract}

\section{Introduction}
The GERDA (GERmanium Detector Array) experiment \cite{gerda} aims to search for the neutrinoless double-beta (0$\nu \beta \beta$) decay of~$^{76}$Ge. The background level in the Ge-diodes is envisioned to be $<10^{-2}$~cts/(keV$\cdot$·kg$\cdot$·y) in the region of interest (Q$_{\beta \beta}\approx2.0$ MeV) for the first phase of the ex\-pe\-ri\-ment, and $<10^{-3}$~cts/(keV$\cdot$·kg$\cdot$·y) for the second phase. That is more than one, re\-spec\-ti\-ve\-ly more than two orders of magnitude lower than in previous 0$\nu \beta \beta$-decay experiments.

The envisioned sensitivity implies very stringent limits on the radiopurity of a variety of materials used. Particularly, the natural contamination of materials close to the Ge-diodes should not exceed levels in the range of $\sim$1 mBq/kg for $^{232}$Th and $^{238}$U \cite{gerda}. In a few cases activities above $\sim$10~$\mu$Bq/kg are not tolerable. Therefore concentrations of Th and U on levels down to $\sim$2.5$\cdot10^{-12}$~g/g and $\sim$0.8$\cdot10^{-12}$~g/g, respectively, have to be achieved and proven by measurements.

Thus, the requirements on material screening include high-sensitivity low-background detection techniques. The most direct way to assess the contamination, that could create dangerous background in the Ge-detector array of GERDA, is to use Ge-spectrometry \cite{Hult0}. The advantage of this screening technique is that the potential sources of $\gamma$-ray background can be directly detected and sensitivities on the order of $\sim$10 $\mu$Bq/kg can be reached \cite{Heusser04}.

\section{Material screening laboratories}
The GERDA collaboration includes several laboratories in which material screening is done, located in four countries across Europe. 
Besides $\gamma$-ray spectrometry using Ge-detectors, also other high-sensitivity techniques for contamination measurements are used in these laboratories. These include low-background proportional gas counters, mass-spectrometry and neutron activation analysis (NAA). An overview of the material-screening laboratories is shown below and summarized in Table \ref{tab:labs}.

\begin{itemize}
	\item[\bf 1.] {\bf Max-Planck-Institut f\"ur Kernphysik (MPIK), Heidelberg, Germany}

The low-background laboratory is located in a shallow underground with 15~m of water equivalent (m~w.e.) shielding overhead. It contains four low-background Ge-spec\-tro\-me\-ters, each with a coaxial p-type crystal, N$_{2}$-flushed sample chamber, active $\mu$-veto system and a $\sim$20~cm thick shield consisting of Pb, Fe and Cu layers (for description of the newest detector, see the second part of this article \cite{Mark}). The efficiency determination is done using the Geant4-based \cite{g4} Monte Carlo tool MaGe \cite{MaGe} (Geant4 version 8.2p01), with $\gamma$-line intensities from the Geant4 default decay database ENSDF \cite{ENSDF}.

\parskip0cm
\parindent0.5cm Two of the detectors can be used for large samples (20~cm x 20~cm x 27~cm). The best obtained sensitivities reach down to 1~mBq/kg \cite{Heusser91}. In case that a material with even better radiopurity is required, the spectrometers can be used for a preliminary selection of candidate samples.

\parskip0cm
Measurements of Rn concentrations in gases or liquids, and Rn emanation from surfaces of materials are performed using proportional gas counters \cite{Grzegorz}. The emanation technique reaches sensitivities of $\sim$1 $\mu$Bq/m$^{2}$ (corresponds to detecting a few tens of Rn atoms). Additionally, mass-spectrometry of noble gases is also performed at MPIK \cite{Grzegorz}.

	\item[\bf 2.] {\bf Laboratori Nazionali del Gran Sasso (LNGS), Assergi, Italy}

\parindent0cm The LNGS deep underground laboratory is located at a depth of 3800 m w.e. Reduction of the $\mu$-flux at this level is $\sim$10$^{-6}$ with respect to the surface. A material-screening facility inside the laboratory includes 10 Ge-detectors for low-level $\gamma$-ray measurements. Detection efficiency is determined using Geant4.7.0 with ENSDF decay database.

\parskip0cm
\parindent0.5cm The most sensitive spectrometer is the GeMPI, which can measure activities on the level of $\sim$10 $\mu$Bq/kg \cite{GEMPI2000}. Currently, two additional detectors of the same design have been constructed at LNGS. They feature $\sim$2 kg coaxial p-type crystals, with large N$_{2}$-flushed sample chambers and Pb/Cu passive shields, all located inside air-tight housing (detailed description of the design and construction of the GeMPI-3 spectrometer can be found in the second part of this article \cite{Mark}).

\parskip0cm Additional screening techniques employed at LNGS include proportional gas counting and mass-spectrometry. NAA was also performed for GERDA, with samples irradiated at Laboratorio Energia Nucleare Applicata (LENA) in Pavia, Italy.

	\item[\bf 3.] {\bf Institute for Reference Materials and Measurements (IRMM), Geel, Belgium}

\parindent0cm IRMM operates seven low-background HPGe-spectrometers in the HADES underground la\-bo\-ra\-to\-ry (500~m~w.e.) \cite{HADES}, which is located on the site of the Belgian nuclear centre SCK$\bullet$CEN in Mol. Amongst the spectrometers are both n-type crystals featuring sub-micron dead layers as well as p-type crystals with $\sim$0.7~mm dead layers. Two of the detectors are set up in a sandwich configuration to increase the efficiency, which is particularly useful for small samples. All shields are made up of an inner lining of freshly produced electrolytic copper (3~to~14~cm), layers of old lead (3~to~5~cm) and less-pure lead (10~to~12~cm). Some of the detectors employ plastic scintillators as muon shields.

\parskip0cm
\parindent0.5cm 
Detector Ge-3, listed in Table~\ref{tab:labs}, has a lower background count rate compared to the publication in 2000 \cite{Hult} due to improved surface cleaning of the copper shield, decay of activation products and replacement of the FET which was a source of $^{137}$Cs. 

\parskip0cm
The efficiency determination is based on measuring activity standards. For samples of irregular shape, size and composition, the EGS4 Monte Carlo code \cite{EGS4} is employed to calculate the efficiency transfer factor as well as the coincidence summing corrections. The nuclear decay data are mainly taken from the DDEP (Decay Data Evaluation Project) database \cite{DDEP}.

\parskip0cm
The proximity of HADES to the BR-1 reactor of SCK$\bullet$CEN allows to perform sensitive NAA, especially with the use of short-lived isotopes. This option was employed to determine the enrichment level of GeO$_{2}$ for GERDA, but can be applied for low-level material screening as well.

	\item[\bf 4.] {\bf Joint Institute for Nuclear Research (JINR), Dubna, Russia}

\parindent0.cm The JINR's above-ground laboratory is operating an n-type coaxial crystal with a passive shield consisting of 10 cm of Cu, 10 cm of Pb and 10 cm of B-loaded polyethylene \cite{Dubna}. Detection efficiency calculation is done with Geant4.7.0 with ENSDF decay database. The detector is used for preliminary selection of samples.

	\item[\bf 5.] {\bf Baksan Neutrino Observatory (BNO), Baksan Valley, Russia}

The Baksan underground laboratory of INR RAS, includes a material-screening facility located at a depth of 660~m~w.e. and a new Deep Underground Low-Background Laboratory located at 4900 m w.e. In the shallow laboratory, the IGEX/Baksan HPGe Set-up \cite{Baksan} is installed, which was previously used in the IGEX 0$\nu \beta \beta$-decay experiment \cite{IGEX}. It can o\-pe\-ra\-te with up to four crystals, placed inside 40 cm thick Cu/Pb/PE shield with nitrogen-gas flushing and a liquid-scintillator muon-veto. The setup is situated inside a room with additional 1 m thick concrete/dunite/steel walls. The shallow laboratory also includes a NaI $\gamma$-ray spectrometer. A new Ge-detector dedicated to material screening will be installed in the deep laboratory.
\end{itemize}

\begin{table}[h]
\begin{center}
        \caption{\rm Selected spectrometers of GERDA laboratories.}
        \label{tab:labs}
	\vspace{1ex}
\begin{tabular}{ c  c  c  c  c }
\hline
\hline
 & & \multicolumn{3}{c}{\bf Most sensitive material-screening detector}\\ \cline{3-5}
 \bf Institute & \bf Depth & & & \bf Background \\
 & & \raisebox{1.5ex}[-1.5ex]{\bf Description} & \raisebox{1.5ex}[-1.5ex]{\bf Active mass} & \bf (40-2700) keV \\ \hline
\hline
 & & p-type Ge with active & & \\
\raisebox{1.5ex}[-1.5ex]{MPIK} & \raisebox{1.5ex}[-1.5ex]{15 m w.e.} & and passive shield & \raisebox{1.5ex}[-1.5ex]{0.84 kg} & \raisebox{1.5ex}[-1.5ex]{1280 cts/d} \\ \hline
 & & p-type Ge with & & \\
\raisebox{1.5ex}[-1.5ex]{LNGS} & \raisebox{1.5ex}[-1.5ex]{3800 m w.e.} & passive shield & \raisebox{1.5ex}[-1.5ex]{2.15 kg} & \raisebox{1.5ex}[-1.5ex]{65 cts/d} \\ \hline
 & & p-type Ge with & & \\
\raisebox{1.5ex}[-1.5ex]{IRMM} & \raisebox{1.5ex}[-1.5ex]{500 m w.e.} & passive shield & \raisebox{1.5ex}[-1.5ex]{1.34 kg} & \raisebox{1.5ex}[-1.5ex]{299 cts/d} \\ \hline
 & & p-type Ge with & & \\
\raisebox{1.5ex}[-1.5ex]{JINR} & \raisebox{1.5ex}[-1.5ex]{surface} & passive shield & \raisebox{1.5ex}[-1.5ex]{1.36 kg} & \raisebox{1.5ex}[-1.5ex]{2.55$\cdot·10^{5}$ cts/d} \\ \hline
 & & p-type Ge with active & & \\
\raisebox{1.5ex}[-1.5ex]{BNO} & \raisebox{1.5ex}[-1.5ex]{660 m w.e.} & and passive shield & \raisebox{1.5ex}[-1.5ex]{4 x 1 kg} & \raisebox{1.5ex}[-1.5ex]{250 cts/d} \\ \hline
\hline
       \end{tabular} 
        \end{center}
\end{table}

Selected results of the  $\gamma$-ray  screening measurements done for GERDA are shown in Table~\ref{tab:results}. In terms of radiopurity, they represent some of the cleanest metals and plastics ever measured. These measurements were done with the GeMPI spectrometer at LNGS, which is the most sensitive detector used for routine material screening within the GERDA collaboration and worldwide. Another recent result of the Ge-spectrometry in the frame of GERDA is a survey of activities in a range of steel batches \cite{Werner}.

\begin{table}[ht]
\begin{center}
	\caption{\rm \small Selected samples measured at LNGS with the GeMPI detector. The iron was produced from a First World War battle ship, and was used for many years in various low-level projects. The copper is electrolytic of NOSV quality from Norddeutsche Affinerie AG. Out of the lead samples, the DowRun quality is supplied by JL Goslar GmbH, and the ancient lead is made from roman relics. The PTFE is Dyneon TF~1620, sintered on special request in a clean way by ElringKlinger Kunstofftechnik GmbH.}
	\label{tab:results}
	\vspace{1ex}
		\begin{tabular}{ c  c  c  c  c  c }
\hline
& \multicolumn{5}{c}{\bf Specific activity [mBq/kg]}\\ \cline{2-6}
\raisebox{1.5ex}{\bf Sample} & \bf $^{40}$K & \bf $^{60}$Co & \bf $^{210}$Pb & \bf $^{226}$Ra & \bf $^{228}$Th\\
\hline
\hline
Old ships iron & 1 $\pm$ 0.4 & $\leq$0.018 & - & 0.15 $\pm$ 0.04 & 0.46 $\pm$ 0.14\\
Copper & $\leq$0.088 & $\leq$0.010 & - & $\leq$0.016 & $\leq$0.012\\
Lead DowRun & 0.44 $\pm$ 0.14 & 0.18 $\pm$ 0.02 & 27000 $\pm$ 4000 & $\leq$0.029 & $\leq$0.022\\
Ancient lead & $\leq$0.27 & $\leq$0.025 & $\leq$1300 & $\leq$0.045 & $\leq$0.072\\
PTFE & 0.54 $\pm$ 0.11 & - & - & 0.021 $\pm$ 0.009 & 0.023 $\pm$ 0.015\\
\hline
		\end{tabular}
        \end{center}
\end{table}

\section{Comparison of evaluation accuracy}

To check the reliability and compatibility of results, a test of measurement and evaluation abilities of the GERDA material-screening laboratories was performed. This intercomparison was based on the Environmental Radioactivity Comparison Exercise 2005, organised by National Physical Laboratory of UK \cite{NPL}. The laboratories measured samples containing $\gamma$-emitting radionuclides with activities on 1-20~Bq/kg level. None of the participants had a prior knowledge of the radioisotopic content and activity of the sample. The results were compared to the re\-fe\-rence activities provided by NPL.

All of the involved GERDA laboratories correctly identified all 10 isotopes present in the sample without any false identification. From the measured activities of individual isotopes, relative deviations with respect to the reference values were calculated. The average results from each laboratory are listed in Table \ref{tab:deviations}. A complete report of the comparison of GERDA laboratories can be found in \cite{LRT06}.

\begin{table}[ht]
\begin{center}
        \caption{\rm Mean results of a comparison of GERDA laboratories.}
        \label{tab:deviations}
	\vspace{1ex}
\begin{tabular}{ cr@{ }c@{ }l }
\hline
Institute & \multicolumn{3}{c}{Mean deviation} \\
\hline
MPIK & $-19.2\%$ &$\pm$& 6.5\% \\
LNGS & $-2.0\%$ &$\pm$& 6.8\% \\
IRMM & $1.7\%$ &$\pm$&1.6\% \\
JINR & $-19.9\%$ &$\pm$& 3.5\% \\
BNO & $-17.6\%$ & $\pm$& 3.4\% \\
\hline
       \end{tabular} 
        \end{center}
\end{table}

The results from the LNGS and IRMM are in excellent agreement with the reference values. The deviations of the other laboratories are satisfactory for the purpose of ultra-low background material selection, which puts more importance on sensitivity and reliability. Nevertheless, as a consequence of this intercomparison, a comprehensive effort was initiated to improve the accuracy of MC-based measurement evaluation at MPIK.

\section{Conclusions}
Gamma-ray Ge-spectrometry is the most direct material screening method relevant for critical backgrounds of the GERDA experiment. The GERDA collaboration has several low-level la\-bo\-ra\-to\-ries available, which provide reliable material selection with sufficient capacity to screen the construction materials. The low-background screening techniques used in these laboratories meet the requirements of GERDA to reliably identify background contaminations, and reach the world's best sensitivity.


\section{Acknowledgements}
This work was partially supported by DFG within the SFB Transregio 27 ``Neutrinos and Beyond'' and by INTAS (Project Nr. 05-1000008-7996). The Baksan group acknowledges the support from RFBR (grant 06-02-01050).

\end{document}